\documentclass{osa-article}

\journal{oe}

\articletype{Research Article}

\begin{document}

\title{Line-field Coherent Sensing with LED Illumination}

\author{Celalettin Yurdakul\authormark{1}, David S. Millar\authormark{2}, Hassan Mansour\authormark{2}, Pu Wang\authormark{2}, Keisuke Kojima\authormark{2}, Toshiaki Koike-Akino\authormark{2}, Kieran Parsons\authormark{2,*} and Philip V. Orlik\authormark{2}}

\address{\authormark{1}Boston University, Boston, Massachusetts, USA\\
\authormark{2}Mitsubishi Electric Research Laboratories (MERL), Cambridge, Massachusetts, USA}

\email{\authormark{*}parsons@merl.com} %

\begin{abstract}
We describe a method of low-coherence interferometry based optical profilometry using standard light-emitting diode (LED) illumination and complementary metal-oxide-semiconductor (CMOS) image sensors. 
A line-field illumination strategy allows for the simultaneous measurement of many points in space. Micron scale accuracy and resolution are achieved and demonstrated using a variety of targets.
\end{abstract}

\section{Introduction}

Optical coherence tomography (OCT) is an interferometric imaging technique that coherently detects the optical signal from the sample with a reference signal. 
OCT offers non-invasive, non-contact label-free imaging of specimens with micron scale high-resolution in all three dimensions (3D) \cite{ultra_high_resolution_OCT,submicrometer_axial_resolution,object_profiling_FMCW,high_resolution_ofdr}. 
Owing to significantly improved sensitivity and speed improvement in Fourier domain OCT, OCT has been a vital imaging platform for 3D \textit{in vivo} studies, particularly in clinical practices \cite{MedicalOCT,OCT_Eye,Optical_Biopsy_OCT}. 
The majority of the OCT studies have been conducted in the near-infrared (NIR) regime ranging from $800$ to $1500$~nm due to long penetration depth in eye and skin. 
However, the NIR wavelength regime requires expensive light sources such as superluminescent diodes and swept-source lasers. 
Conversely, the visible window could provide superior lateral resolution while maintaining high axial resolution with narrower relative bandwidth requirements compared with the NIR counterparts. 
The use of narrowband light sources may also reduce the effects of chromatic dispersion. 
Also, the scattering cross-section in the visible window is much larger compared with the longer wavelengths due to the inverse relationship with the fourth power of the illumination wavelength. 
Although this limits the penetration depth in a scattering medium, the visible light could be utilized for shallow depth-range applications such as oxygen saturation measurements \cite{visOCTReview,visOCTJY1,visOCTJY2} and surface inspection of printed circuits \cite{PCBvisOCT}.

Spectral-domain (SD) OCT enables highly sensitive and high-speed axial scans \cite{spectral_domain_OCT}. 
A low-coherence (broadband) light source illuminates the sample, and the spectral data corresponding to the interferometric signal from the sample and reference arms is recorded by a spectrometer. 
This interference spectrum is later used to reconstruct the sample's depth information via Fourier transformation. 
Linescan cameras have been employed in SD-OCT setups where a focused spot needs to be scanned over the sample. 
The line measurement from each spot corresponds to the Fourier transform of an A-scan in a certain depth range limited by the number of pixels in the spectrometer sensor and the coherence length of the light source. 
Owing to the increased prevalence of image sensors in smartphones and other high-volume consumer electronic devices, a complementary metal-oxide-semiconductor (CMOS) camera technology has evolved rapidly, leading to extremely high-performance sensors at remarkably low cost. 
As an alternative and faster approach, a parallel SD-OCT has been introduced in \cite{firstLinefieldOCT}. 
Instead of recording A-scans point by point, multiple A-scans along a line are captured using an area-scan camera, enabling B-scans at a single shot. 
The illumination needs to be scanned in only one direction compared with the conventional SD-OCT. 
Moreover, the diffraction-limited lateral resolution along the line-field can be achieved. 
Although parallel SD-OCT has been successfully demonstrated in applications spanning from biology \cite{LFSDOCTBioSensitivity,LFSDOCTBiofast,LFSDOCTBio1,LFSDOCTBio2,LFSDOCTBio3} to optical meteorology \cite{LFSDOCT1, LFSDOCTAutomative, LFSDOCTLens,LFSDOCTthinFilm}, they typically rely on expensive light sources such as supercontinuum lasers and superluminescent diodes. 
Here, we report a fast and low-cost CMOS camera based line-field SD-OCT technique with visible light-emitting diode (LED) illumination for coherent depth imaging. 
We utilize an off-the-shelf green LED and a machine vision camera. 
Our technique provides a high-resolution, low-cost, non-destructive, and contact-free 3D imaging tool for industrial applications. 

\begin{figure}[tb]
\centering\includegraphics[width=\linewidth]{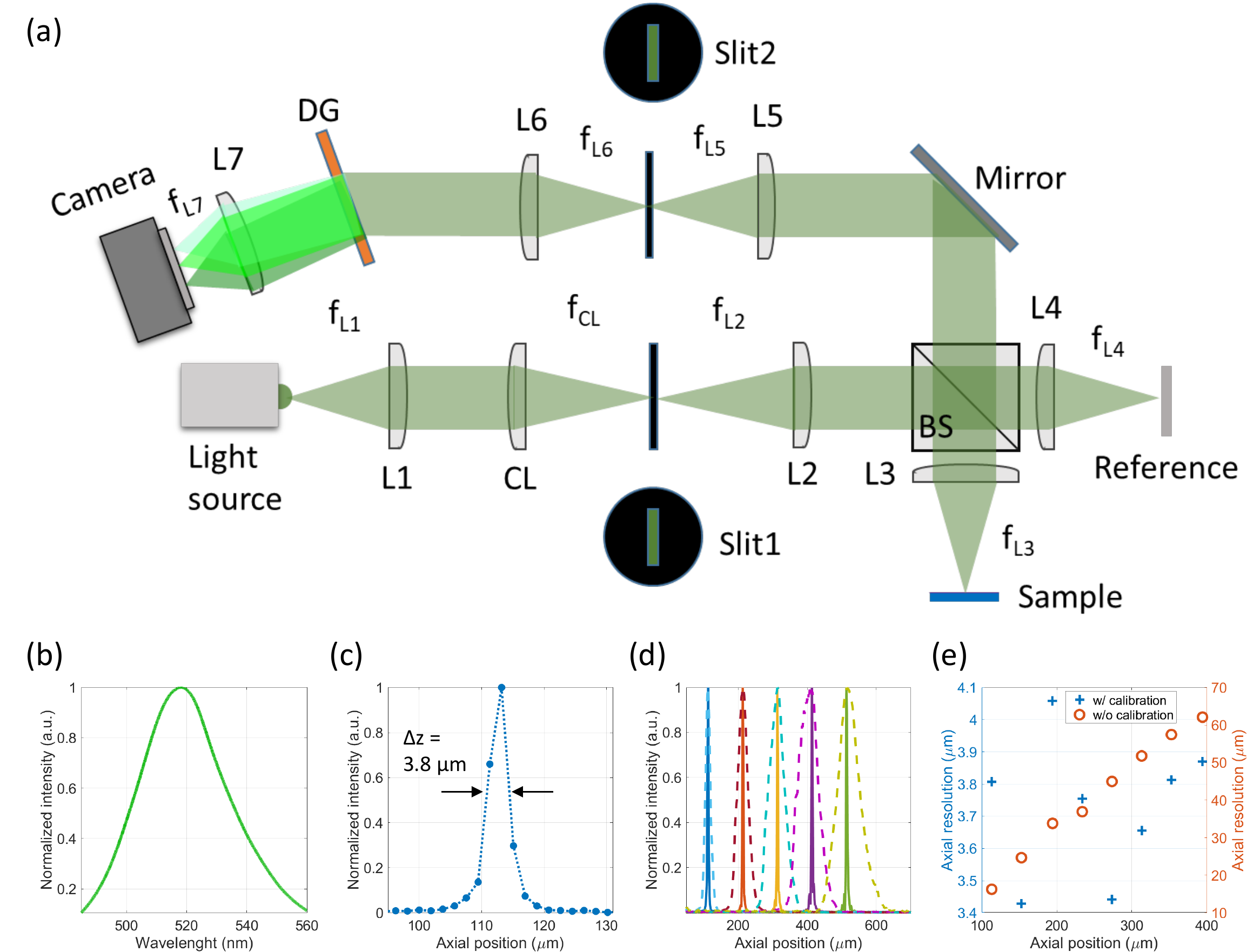}
\caption{(a) Schematic of the visible line-field depth sensor setup. 
L1--L7; achromatic doublets; CL: cylindrical lens; BS: beam splitter; DG; diffraction grating. 
(b) LED light source spectrum recorded at the camera. 
(c) Wavenumber linearized axial resolution ($\Delta z = 3.8$~µm). 
The curve is calculated from the interferogram measured at the axial distance of $115$~µm. 
(d) Axial responses at various depth positions, both with and without wavenumber linearization. 
(e) The calculated axial resolutions at given range in (d). 
The resolutions are obtained from FWHM calculations.}
\label{fig:fig1}
\end{figure}

\section{Methods}
\label{sec:methods}
\subsection{Experimental setup}

The schematic of the experimental setup is shown in Fig.~\ref{fig:fig1}. 
A partially coherent narrowband green LED (Thorlabs, M530L4) with $\lambda = 530$~nm central wavelength and $\Delta \lambda = 35$~nm bandwidth is employed as a light source. 
The calculated coherence length of the LED is $l_c = 7$~µm, which can provide about $3.5$~µm axial resolution. 
The light emitted from the LED is collimated by an aspheric condenser (L1 with a focal length of $f=25$~mm) with a diffuse surface. 
This lens reduces the LED chip image that creates heterogeneous illumination patterns in the far field. 
A bull's eye apodization filter is placed after the collimator lens to achieve a uniform beam shape which directly affects the line-field illumination uniformity. 
A cylindrical lens (CL, $f=50$~mm) asymmetrically focuses the collimated light on an adjustable mechanical slit (Slit 1). 
The pairs of identical achromatic doublets, L2--L3 and L2--L4, image the Slit 1 onto the sample and reference planes through a $50:50$ unpolarized beam splitter to create line-field illumination. 
In other words, the slit is in the conjugate plane of the sample and reference planes. 
The nominal slit width is set to $10$~µm, which is close to the camera pixel limited resolution of $9$~µm. 

A highly reflective silver mirror is mounted to provide a reference signal for coherent detection. 
The line-fields on both reference and sample planes are relayed onto a spatial filter using $1:1$ relay pairs of L3--L5 and L4--L5. 
The relayed line field is spatially filtered by the Slit 2 where its width is set to Slit 1's width size $10$~µm. 
L6 collimates the spatially filtered light. 
The collimated beam illuminates the transmission grating (Wasatch Photonics, $1800$ lines/mm $@$ $532$~nm) at about $28^\circ$ that maximizes diffraction efficiency. 
The angle of incidence on the grating is empirically determined by maximizing the peak power on the camera. 
The imaging lens (L7) focuses the diffracted light onto the CMOS camera sensor (FLIR, BFS-U3-70S7M-C) which has $2200 \times 3208$ pixels. 
Each row captures interference spectrum of a point on the line-field illumination. 
Therefore, multiple interference spectra are recorded simultaneously. 
This parallel detection scheme enables a less complex experimental setup and improves the throughput of the system. 
The camera angle and height are adjusted to optimize the collection efficiency. 

Note that L2--L7 are all identical achromatic doublets with $100$~mm focal distances. 
Overall, our imaging system has $1 \times$ magnification and pixel size limited lateral resolution of $9$~µm. 
The field-of-view (FOV) provided by the line-field illumination is $9.9$~mm. 
The sample is scanned along the lateral direction by a motorized XY stage (Zaber, ASR100B120B-T3A). 
The image acquisition and stage control are automated by a custom-built Python software. 
The nominal wavelength sampling resolution of our system is $17$~pm. 
The theoretical depth-range, limited by Nyquist theorem, is $12.1$~mm but we limit our study to $1$~mm axial range due to the signal-to-noise ratio (SNR) drop.

\subsection{Wavenumber calibration}
Non-linear sampling of $k$-space significantly deteriorates axial point spread function (PSF)---particularly at longer distances---broadening the axial response at longer distances and sensitivity of the system \cite{calibrationSNRres}. 
This comes from the fact that the spectrometer samples the light linearly in the wavelength ($\lambda$) space. 
There have been various methods developed in the OCT literature to overcome this nonlinear sampling of the $k$-space \cite{spectralCalibrationOSA,spectrometerCalibration,wavenumberCalibrationCO}. 
The common and expensive solution is to calibrate the system with a known narrow linewidth light source, or to calibrate against another instrument with a known response such as an optical spectrum analyzer \cite{spectralCalibrationOSA}. 
However, these methods require either well-developed light sources at a particular wavelength band of interest or expensive instruments. 

Here, we implemented a simple SD-OCT calibration method using the phase of the Hilbert transformed interference spectra acquired at known axial displacements, introduced by Wang \textit{et al.} \cite{wavenumberCalibrationCO}. 
The calibration procedure simply relies on the measurement of linear phase accumulation with respect to axial position $z$. 
The OCT signal captures the coherence sum of reflected field $E_r$ and signal field $E_s$. 
The intensity signal captured at the camera for a given frequency $ k = \frac{2\pi}{\lambda}$ can be expressed as follows:
\begin{equation}
    I_\mathrm{detector}(z) = |E_r|^2 +|E_s|^2 + 2|E_r||E_s| \cos(2k\Delta z),
    \label{eq:OCT_intensity}
\end{equation}
where $\lambda$ is the illumination wavelength and $z$ is the sample's axial position with respect reference arm. 
The first two terms constitute direct component (DC) intensity contributions from the reference and sample, respectively. 
The sample's depth information is encoded into the modulation frequency of the interferometric term. 
After filtering out the DC part, the signal of interest for axial response calculations becomes:
\begin{equation}
    I_m(z) = 2|E_r||E_s| \cos(2k\Delta z).
    \label{eq:OCT_crossterm}
\end{equation}
It is clearly seen from the equation above that linear wavelength sampling leads to non-linear $k$-space sampling. 
To linearize the recorded spectral interferograms in $k$-space, we simply obtain two image sets $\Delta z_1 - \Delta z_2 = \Delta z_c$ apart from each other. 
The phase difference between two signals becomes $\Phi = 2k\Delta z_c$ which is a linear function of frequency $k$. 
In theory, any linear measurement which is a function of $k$ such that $y_n = g(k_n)$ where $g$ is a linear function at any domain can be utilized to linearize the $k$-space \cite{spectrometerCalibration}. 

We calculate the phase of each image simply by taking Hilbert transforms followed by phase unwrapping. 
The phase difference between two measurements realizes the non-linear phase curve function, which is expected to be linear in the ideal case of $k$-space spectrometer. 
Then, this phase curve is used for interference spectrum interpolation in linear $k$-space representations. 
To improve the accuracy of interpolation \cite{zeroPaddingAccuracy}, the captured spectra images are zero-padded in the frequency domain by simply taking a $2^{14}$ point inverse fast-Fourier transform (IFFT), which increases spectral resolution by nearly five-fold. 

\begin{figure}[tb]
\centering\includegraphics[width=\linewidth]{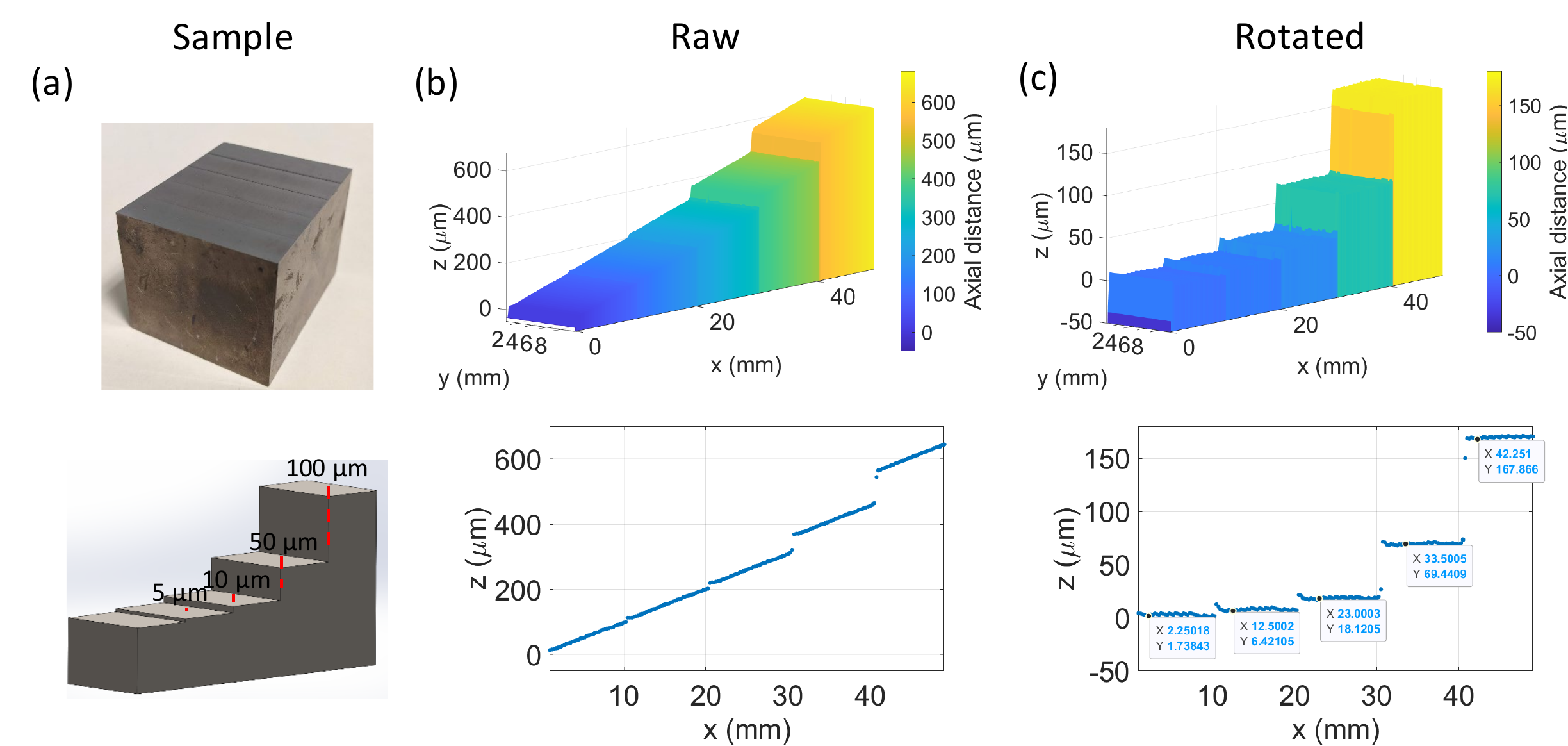}
\caption{Surface topography of test target. 
(a) Top: The picture; Bottom: 3D schematic of the stair test target. 
(b) Top: Surface topography; Bottom: cross-section profile from the raw measurement. 
(c) Top: Surface topography; Bottom: cross-section profile obtained after correcting for the tilt along the sample scan axis. 
A pre-calibrated 3D rotation matrix is applied for the correction. 
(d) The cross-section profiles are obtained by averaging along the $y$ axis.}
\label{fig:fig2}
\end{figure} 

\subsection{Line-field image reconstruction framework}
The signal spectrum in the captured intensity images is in the form of Eq.~(\ref{eq:OCT_intensity}). 
The measured spectrum has a spectral fringe pattern on top of the DC spectra from sample and reference signals. 
The fringe frequency is defined by the optical path length difference between sample and reference arm as shown in Eq.~(\ref{eq:OCT_crossterm}). 
The recorded raw spectra images are high-pass filtered, followed by the $k$-space linearization process explained above. 
The filtered DC signal's amplitude is saved for reflectance calculations. 
Finally, the depth information from the sample is reconstructed from the $k$-space linearized interference spectra using computationally efficient fast-Fourier transform (FFT) functions. 
The frequency domain of the interference spectra is low-pass filtered by a Gaussian smoothing kernel with a standard deviation of $3$. 

The local maximum frequency bin of each interference spectrum encodes the axial displacement of its corresponding point on the line-field.  
The maximum searching algorithm is applied to extract the index position. 
The index position is then converted into the nominal axial displacement. 
To calculate the conversion ratio, we acquire a series of calibration measurements by scanning a gold mirror sample at a known step size of tens of microns. 
The maximum indices of these measurements are extracted to calculate by dividing the known displacement between two measurements by their index difference. 
The conversion ratios calculated from multiple measurement pairs are averaged to increase the calibration accuracy.

\section{Results}

\subsection{System characterization}
The custom-built spectrometer in our setup provides quasi-linear sampling of wavelength space. 
This is a valid assumption under the given narrowband light source $\Delta\lambda = 35$~nm and diffraction angle of $\Delta\theta \simeq 6 ^\circ$. 
We implement the calibration method described in Section~\ref{sec:methods}. 
Fig.~\hyperref[fig:fig1]{1b} shows the spectrum of the LED measured by our system. 
The full-width-half-maximum (FWHM) of axial PSF calculated at $115$~µm distance is found as $\Delta z = 3.8$~µm. 
Our experimental finding is consistent with the theoretical axial resolution of $3.5$~µm. 
As shown in Fig.~\hyperref[fig:fig1]{1d}, the axial PSFs broaden in the absence of the calibration and therefore drastically limits the axial resolution. To further emphasize the resolution improvement, we show FWHM calculated at various depths (see Fig.~\hyperref[fig:fig1]{1e}). 
While the PSF without the calibration is significantly broadening from $16$ to $62$ microns, the calibrated PSF is nearly constant around $3.8$~µm with a standard deviation of $0.2$~µm.

\subsection{Surface topography of metallic surfaces}
To experimentally demonstrate our system's functionality, we first image specularly reflective samples, including precision machined steel test target and coins. 
To characterize the system performance and resolution limit, we show the high-precision steel test target image in Fig.~\ref{fig:fig6}. 
The test target consists of four steps: (i) $5$~µm, (ii) $10$~µm, (iii) $50$~µm, and (iv) $100$~µm. 
To capture the surface topography, we scanned the line-field from one end to other end along center of the target (Fig.~\hyperref[fig:fig2]{2a}). 
We observed about $500$~µm axial shift over a $50$~mm lateral scan range due to mechanical tilt on the sample holder (see Fig.~\hyperref[fig:fig2]{2b}). 
We calculate the tilt angle along the scanning axis as less than $10$~mrad. 
To prevent the data misinterpretation and correct for the tilt, the calculated surface topography is rotated around the $x$ axis using 3D rotation matrices. 
As shown in Fig.~\hyperref[fig:fig2]{2b-c}, the rotated cross-section profile is significantly improved compared with that of the raw. 
This rotation matrix is applied to all line-field images. 
We also emphasize that our system can distinguish and resolve $5$~µm step which is close to theoretical axial resolution limit of $3.5$~µm.

\begin{figure}[t]
\centering\includegraphics[width=\linewidth]{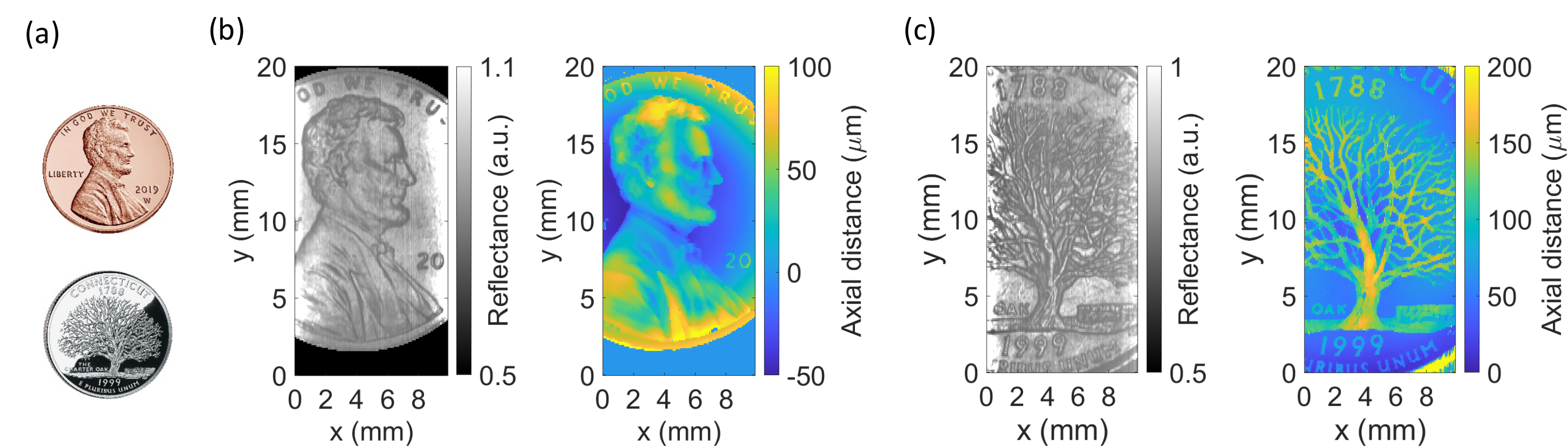}
\caption{(a) Pictures of one-cent Lincoln and twenty-five cent Connecticut coins. 
(b) Left: Reflectance; Right: surface topography images of Lincoln embossment obtained from the visible line-field depth sensor. 
(b) Left: Reflectance; Right: surface topography images of Oak tree embossment obtained from the visible line-field depth sensor. 
The coin images are obtained from the United States Mint website \cite{USMint}.}
\label{fig:fig3}
\end{figure}

Next, we show the topographic reconstruction of the Lincoln and Connecticut coins in Fig.~\ref{fig:fig3}. 
The topographic map of the Lincoln embossment is recovered with high data fidelity. 
Since the structure consists of mostly coarse features, we simply applied a median filter with a $3 \times 20$ window size to remove noise artifacts due to low reflection at the structure edges. 
The total depth range is approximately $150$~µm. 
We also provide the reflectance images calculated from the magnitude of the interferograms as described in Section~\ref{sec:methods}. 
The reflectance nonuniformity around the periphery in $x$ axis stems from the fact that the illumination intensity drops towards the edges across the line-field. 
Although the apodization filter improves the illumination uniformity at the center FOV, the edges still suffer, due to the vignetting. 
This can be further improved by using larger optics (e.g, $2$~in) or shorter line-field illumination. 
Yet, the reconstructed images still have enough SNR to retrieve the depth information as demonstrated in Fig.~\hyperref[fig:fig3]{3c}. 
The tree branches in Connecticut coin is successfully retrieved, indicating a sufficient lateral resolution.

\begin{figure}[t]
\centering\includegraphics[width=\linewidth]{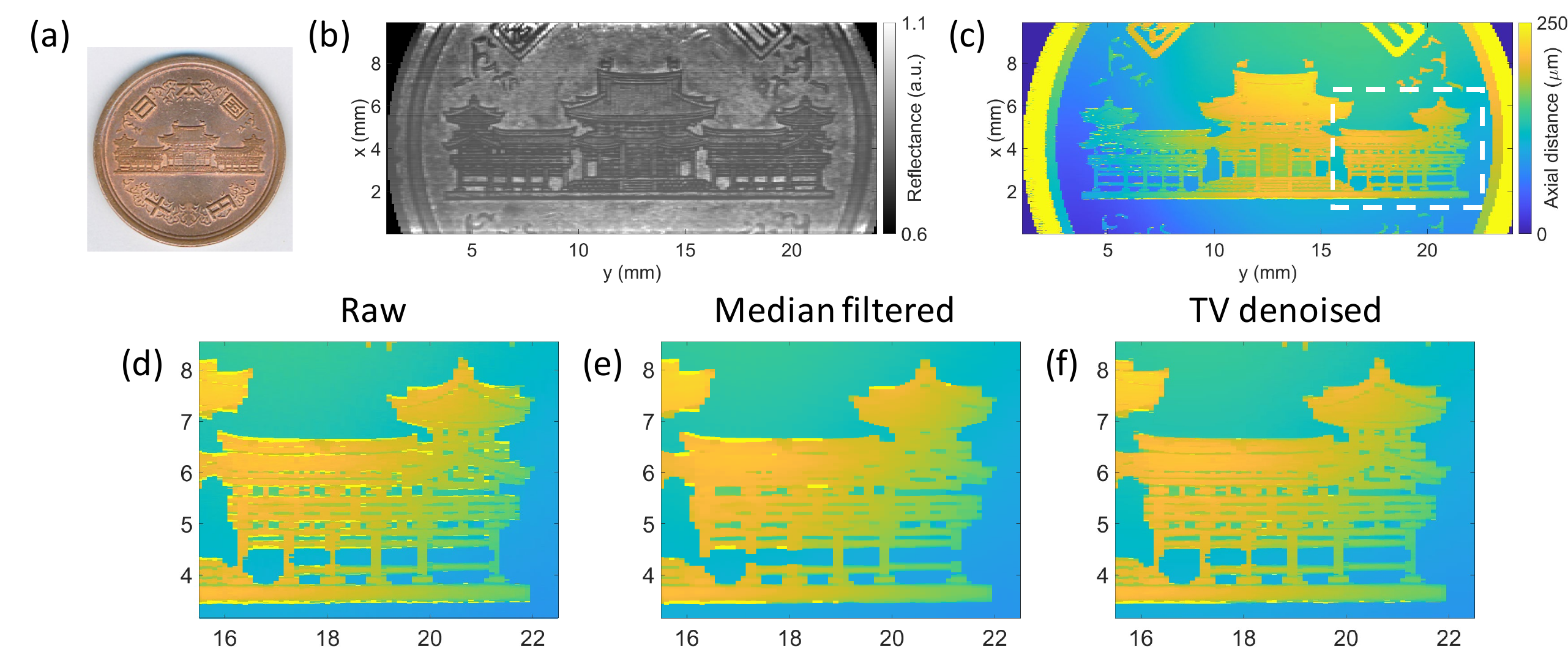}
\caption{(a) Picture of the Byodo-in Temple on the Japanese $10$ Yen coin. 
(b) Reflectance and (c) TV-denoised surface topography images obtained from the visible line-field depth sensor. 
TV regularized image inpainting is applied to remove unwanted depth estimation at low reflection points. 
(d) Raw, (e) median filtered, and (f) TV-denoised zoom-in images indicated by the dashed rectangle in (c). 
TV inpainting preserves high-resolution features within the image.}
\label{fig:fig4}
\end{figure}

\begin{figure}[ht]
\centering\includegraphics[width=.9\linewidth]{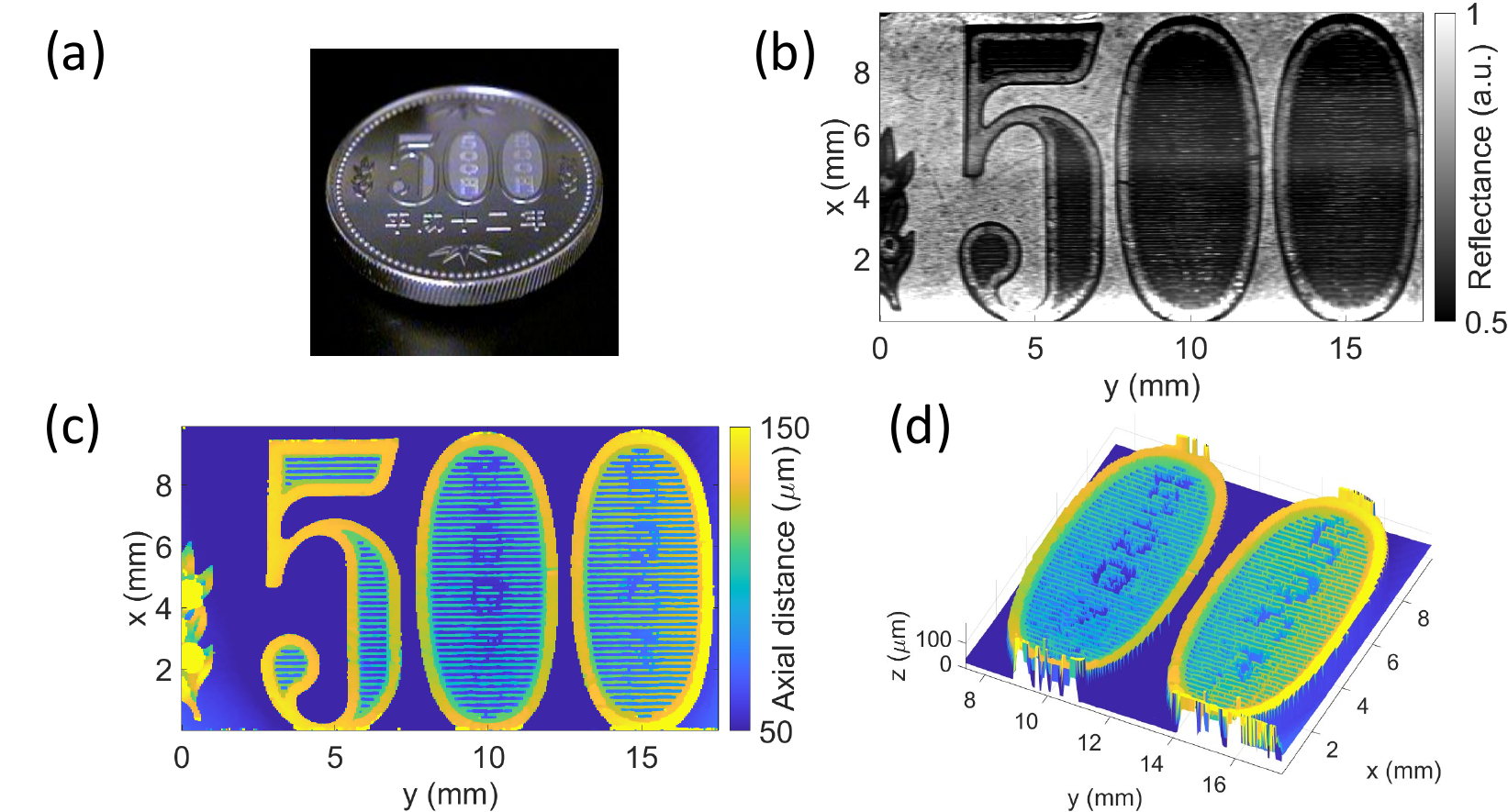}
\caption{(a) Picture of the Japanese $500$ Yen coin. 
(b) Reflectance and (c) surface topography images obtained from the visible line-field depth sensor. 
(d) 3D surface plot that emphasizes visualization of anti-counterfeit hologram at zeros, which can only be seen at an angle. 
The latent characters read $500$ (numerical) yen (Japanese character). 
The coin image is obtained from Japan Mint website \cite{JapanMint}.}
\label{fig:fig5}
\end{figure}

To improve the accuracy of reconstruction at the low-SNR pixels, we implement total-variation (TV) regularized image inpainting that promotes piecewise smoothness in the reconstructed topography map. 
Let $\tilde{X} \in \mathbb{R}^N$ denote the estimated topography map image rearranged into a vector of $N$ pixels. 
The technique first identifies the set of high-SNR pixels $\Omega$ and its complement set of low-SNR pixels $\Omega^c = \{1\cdots N\}/\Omega$. 
A sampling operator $\mathbf{A}: \mathbb{R}^N \rightarrow \Omega$ is constructed to select the high-SNR pixels $Y$ from the estimated topography image $\tilde{X}$, such that
\begin{equation}
    Y = \mathbf{A}\tilde{X}.
\end{equation}

The inpainting technique then fills in the low-SNR pixels on the set $\Omega^c$ by solving the following optimization problem to reconstruct the denoised tomography map $X^*$:

\begin{equation}
    X^* = \arg\min\limits_{X} \frac{1}{2}\|Y - \mathbf{A} X\|^2_2 + \lambda\|X\|_{\text{TV}},
\end{equation}
where 
\begin{align}
    \|X\|_{\text{TV}}=\sum\limits_{i=1}^N \sqrt{(\nabla_x X)^2(i)+(\nabla_y X)^2(i)},
\end{align}
is the isotropic total variation penalty function, $(\nabla_x X)$ and $(\nabla_y X)$ denote the respective horizontal and vertical gradients of the image, and $\lambda$ is a regularization parameter that controls the amount of smoothness encouraged by the TV regularization. 

We demonstrate fine-detailed structure topography using the Japanese $10$ Yen coin in Fig.~\ref{fig:fig4}. 
The coin depicts a temple with sharp and dense features such as windows and pillars. 
As seen from reflectance image in Fig.~\ref{fig:fig4}, these structures have lower reflectivity at the edges. 
The low-reflectivity points reduce the measurement SNR at the corresponding pixels. 
Therefore, the depth recovery at the edges becomes more challenging. 
As shown in the magnified images (Fig.~\hyperref[fig:fig4]{4e-f}), the image inpainting based on TV regularization significantly improves the image quality of the reconstructed surface tomography. 
In contrast, the high-resolution information including densely separated temple windows are lost in the median filtered image. 
Such fine features are preserved in TV-denoised image. 
Hence, applying TV denoising is verified as a viable solution for the low-SNR and detailed samples.

Next, we investigate the capability of the line-field sensor on a $500$ Yen coin which has latent characters that can only be seen at an angled view (Fig.~\ref{fig:fig5}). 
This anti-counterfeit hologram features can be distinguished in the depth maps whereas the reflectance image does not show any structural difference. 
This specular hologram is further visualized by an angled 3D surface plot. 
We speculate that these features are machined at different height recesses compared to the adjacent lines. 
Such features cannot be differentiated by a naked eye or camera since the structures have very small depth differences well below the axial resolution of the conventional 2D imaging systems.

\subsection{Surface topography of diffusive reflective surfaces}

\begin{figure}[tb]
\centering\includegraphics[width=.9\linewidth]{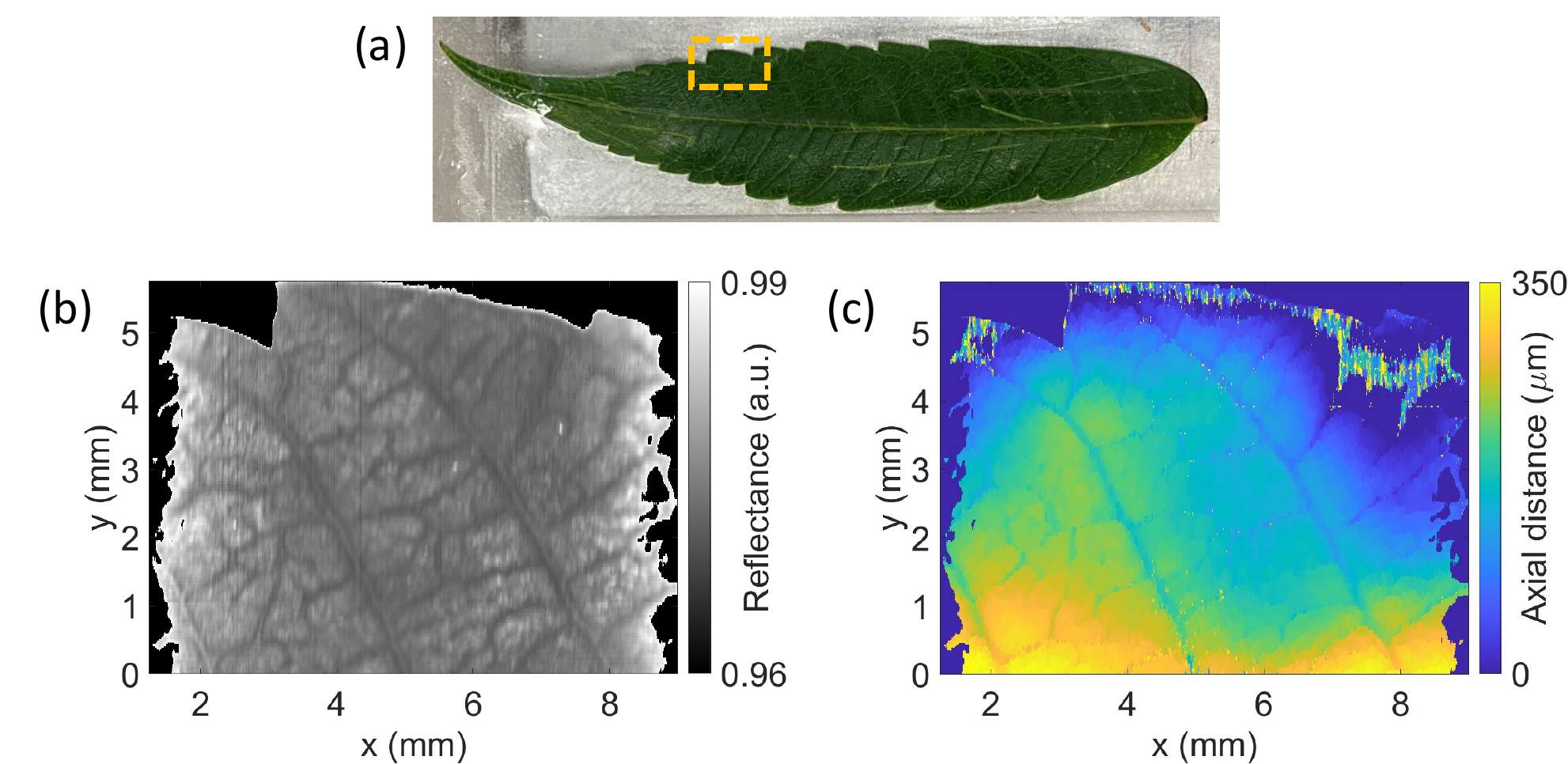}
\caption{Surface topography of leaf sample obtained from the visible line-field depth sensor. 
(a) Picture of the leaf sample. 
(b) Reflectance and (c) surface height images of the area indicated in (a). 
TV-regularized image inpainting is applied to remove unwanted depth estimation particularly at low reflection points.}
\label{fig:fig6}
\end{figure}

To highlight the application of the system under consideration to different types of samples with low and diffusive reflective surfaces, we measured a leaf sample, shown in Fig.~\ref{fig:fig6}. 
The depth profile along the small capillaries are consistent with the reflectance image. 
The estimated total depth range was larger than $400$~µm. 
The height increase towards the leaf center is due to the elevation from thick midrib structure. 
We note that the reflectance of the leaf surface is around $1$--$2$\% compared to the reflectance of the mirror reference. 
The low reflectance from leaf is close to our detection limit of $0.7$\% contrast. 
Since our technique relies on interferometric detection, the sensitivity is limited by shot-noise. 
We calculated the detection limit for saturating the pixels up to $80$\% of their well-capacity. 
This limit can be further improved by using larger well-depth sensors or frame averaging at the cost of speed.

\section{Conclusions}

We have proposed and demonstrated a non-destructive and contact-free method for coherent optical profilometry using LED illumination and CMOS image sensors. 
Micron-scale accuracy and resolution were demonstrated with a variety of targets, including a machined metallic target with a known ground-truth, various coins, and a leaf. 
We note that the method proposed in this work can recover the specular hologram embedded in the Japanese $500$ Yen coin, from an angle at which it is not visible to reflectance only imaging. 
We also note that the fine structures present in the surface of the leaf are reconstructed, despite their weak and diffuse reflection. 
To improve the sensitivity for weakly scattering/reflective samples, high-power LEDs and neutral density filter at the reference arm can be employed. 
Although it is not demonstrated in this study, our line-field profilometer can be applied to transparent samples with multiple reflection boundaries including coatings on glass surfaces and wafers. 

\section*{Funding}

We have no funding to disclose.

\section*{Disclosures}

We have no interests to disclose.

\bibliography{reference}

\end{document}